\documentclass[]{spie}  

 \usepackage{gensymb}
\usepackage{amsmath,amsfonts,amssymb}
\usepackage{graphicx}
\usepackage[colorlinks=true, allcolors=blue]{hyperref}
\usepackage[numbers]{natbib}

\title{CUTE Data Simulator and Reduction Pipeline}

\author[a]{A. G. Sreejith}
\author[a]{Luca Fossati}
\author[a]{Manfred Steller}
\author[b]{Brian T. Fleming}
\author[b]{Kevin France}

\affil[a]{Space Research Institute, Austrian Academy of Sciences, Schmiedlstrasse 6, 8042 Graz, Austria}
\affil[b]{University of Colorado, Laboratory for Atmospheric and Space Physics, UCB 600, Boulder, CO, 80309, USA}

\authorinfo{Further author information: (Send correspondence to A. G. Sreejith)\\A. G. Sreejith: E-mail: sreejith.aickara@oeaw.ac.at}

\begin{document} 
\maketitle

\begin{abstract}
The Colorado Ultraviolet Transit Experiment (CUTE) is a 6U NASA CubeSat carrying a low-resolution (R\,$\approx$\,3000), near-ultraviolet (255 -- 330\,nm) spectrograph fed by a rectangular primary Cassegrain. CUTE, is planned for launch in spring 2020 and it will monitor transiting extra-solar planets to study atmospheric escape. We present here the CUTE data simulator, which is a versatile tool easily adaptable to any other mission performing single-slit spectroscopy and carrying on-board a CCD detector. We complemented the data simulator with a data reduction pipeline capable of performing a rough reduction of the simulated data. This pipeline will then be updated once the final CUTE data reduction pipeline will be fully developed. We further briefly discuss our plans for the development of a CUTE data reduction pipeline. The data simulator will be used to inform the target selection, improve the preliminary signal-to-noise calculator, test the impact on the data of deviations from the nominal instrument characteristics, identify the best spacecraft orientation for the observation of each target and construct synthetic data to train the science team in the data analysis prior to launch.
\end{abstract}


\section{Introduction}\label{sec:intro}
Escape, a process leading atmospheric gas to leave the planet's gravitational well and disperse into space, plays a key role in the long-term evolution of a planetary atmosphere. One of the requirements for a planet to be habitable is to host a secondary atmosphere, which is currently believed to be acquired or generated following the escape of the primary, hydrogen-dominated atmosphere \citep{lammer}. The rapid hydrodynamic loss that has, for example, affected the primary atmospheres of the early Venus, Earth, and Mars \citep{lammer2018} cannot be studied on any object in the solar system, but it can be fortunately observed on close-in extrasolar planets. Atmospheric escape and the interaction of the upper atmosphere with its environment (i.e., stellar wind) can be best observed at ultraviolet (UV) wavelengths. This is because the optical depth of the escaping material is low at longer wavelengths. The first detection of an escaping exoplanet atmosphere was achieved by Vidal-Madjar et al. (2003) \citep{Vidal} who used HST/STIS HI Ly$\alpha$ transit detect the extended upper atmosphere (exosphere) of the hot-Jupiter HD209458b. Since this first detection of an exoplanetary exosphere, UV transmission spectroscopy led to the detection of atmospheric escape for the hot Jupiters WASP-12b and HD189733b and the mildly-irradiated Neptune-mass planet GJ436b \citep{fossati2010,haswell2012,lecavelier2012,kulow2014,ehrenreich2015}. 

The UV exoplanet transit observations conducted so far led to the detection of a large variety of phenomena, such as excess transit depths up to 50\%, and significant transit asymmetries. In addition, some of these phenomena have been found to be highly time variable \citep[e.g.,][]{haswell2012}. However, at present, the theories explaining these phenomena exceed by more than a factor of 40 the number of relevant transit observations. A thorough observational study of these phenomena would require an intensive observational campaign at UV wavelengths, which could be pratically undertaken only by the Hubble Space Telescope (HST), which is our almost only UV ``eye''. Such a large observational effort would require thousands of HST orbits, which are unschedulable on a general-purpose shared-used facility such as HST, more so given its now limited life-time. Thanks to the large size of escaping atmospheres and to the short orbital periods of the planets expected to host an extended atmosphere, atmospheric escape can be studied in detail with a dedicated high-efficiency instrument operating at near-UV wavelengths (250--320\,nm) and attached to a small telescope. 

The Colorado Ultraviolet Transit Experiment (CUTE) is a 6U-form CubeSat that has been specifically designed to deliver the UV transmission spectroscopy observations that are required to advance our understanding of atmospheric escape. The instrument covers the 255--330\,nm spectral region with an effective resolving power around 2500. 
The CUTE mission is designed to provide exactly the kind of spectroscopic observations necessary for advancing the study of exoplanet atmospheric escape. CUTE will look for the presence of absorption features in near-UV transmission spectra characteristic of atoms and molecules inaccessible from the ground, such as Fe{\sc ii}, Mg{\sc ii}, Mg{\sc i}, and OH. The dedicated CUTE mission architecture and scheduling enables to run a dedicated survey to study several of the phenomena characterizing atmospheric escape and their temporal variability. A detailed description of CUTE mission design and hardware is presented in \citep{fleming}.

This paper presents a brief description of the CUTE data simulator and the main requirements of the data reduction pipeline derived from the results of the simulated data. The CUTE data simulator combines a tailor-made suite of IDL functions treating all known physical (e.g., wavelength-dependent planetary transit, stellar spectral energy distribution) and instrumental (e.g., spectral dispersion, spacecraft jitter) effects playing a role in the CUTE data. We have designed the simulator in such away that it can be easily adapted to other small low and medium resolution spectrographs. 
\section{Temperature Tolerance} 
Before starting the development of the data simulator, we derived the expected characteristics of the instrument affecting the resulting images, such as spectral dispersion and spread of the spectrum in the cross-dispersion direction, using the Zemax ray tracing software. All relevant parameters derived from Zemax are input parameters for the simulator. We further employed Zemax to evaluate the effects of temperature variations on these parameters. This gives an idea of how much these parameters may change, thus of how much the resulting images may change, as a result of temperature variations. This is extremely important for the development and training of the data reduction pipeline, which needs to be able to handle data slightly different from those obtained with nominal conditions.

\begin{figure}[h!]
\begin{center}
\includegraphics[width=\textwidth]{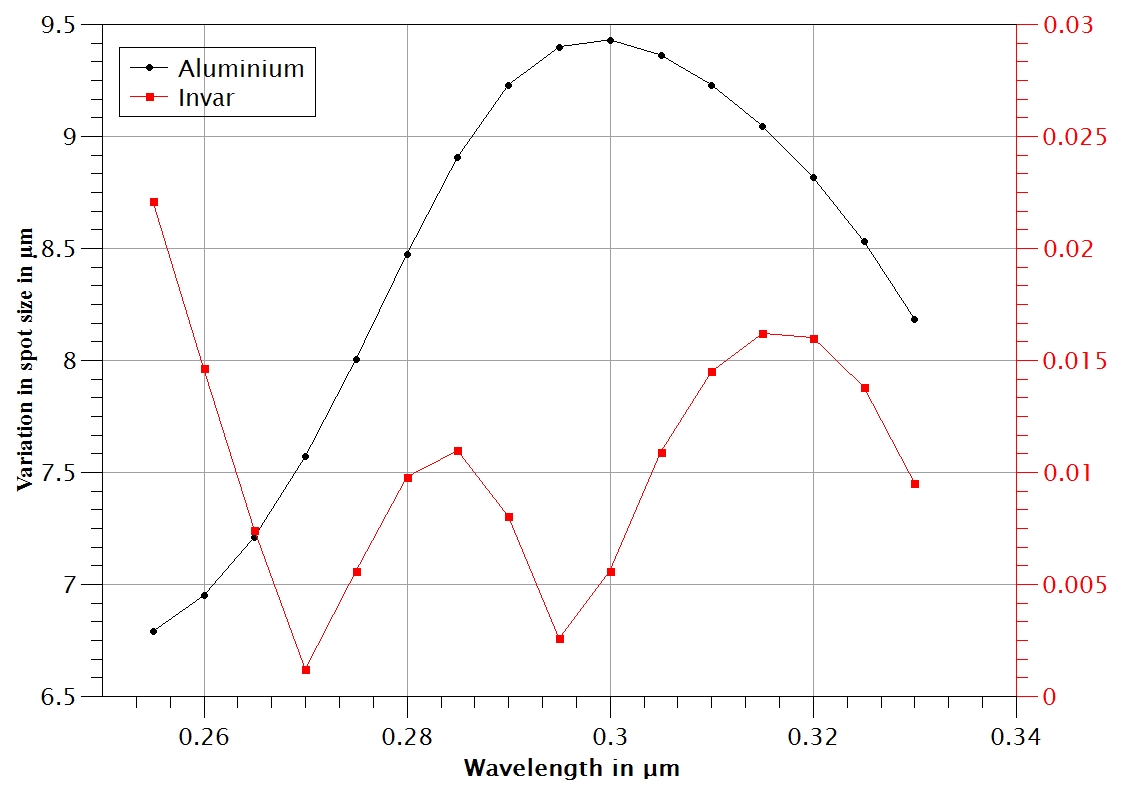}
\caption{Variation in spot size in $\mu$m as a result of varying the temperature between $-$40$\degree$ and $+$40$\degree$\,C for Aluminum (black) and Invar (red). Note the large difference in the scale between the left and right y-axes. CUTE's plate scale is 2.5" per pixel and the size of one pixel is 13.5\,$\mu$m.}
\label{fig:fig1}
\end{center}
\end{figure}

We employed Zemax to carry out a thermal analysis of the entire optical system to find out the magnitude of the variations in the optical performance due to temperature changes assuming different materials. We considered two sets of materials, namely Aluminum and Invar, with temperatures varying from $-40^{\circ}$\,C to $+40^{\circ}$\,C. The results are shown in Figure~\ref{fig:fig1}. We found that the use of aluminum, which has a thermal expansion coefficient of $23.6\times10^{-6}K^{-1}$, will significantly affect the spectral resolution, possibly reducing it by a factor of 1.5--1.7. The use of invar, instead, which has a thermal expansion coefficient of $1.2\times10^{-6}K^{-1}$, does not lead to deviations over one resolution element. A detailed thermal analysis on the entire instrument is being carried out at the Laboratory for Atmospheric and Space Physics, Colorado, USA. 

We have also carried out simulations to look at variations in the instrument performances due to mechanical stresses. We found that manufacturing tolerance of up to 200\,$\mu$m can be corrected by adjusting the focus during assembly and testing before launch. We are planning on having the telescope fixed and staked once focused and aligned. The grating will have shim-only adjustment capability, and the fold mirror will have 5 mm of adjustment in a piston-tip-tilt mount. 
\section{Data Simulator}
\begin{figure}[h!]
\begin{center}
\includegraphics[height=20.5cm]{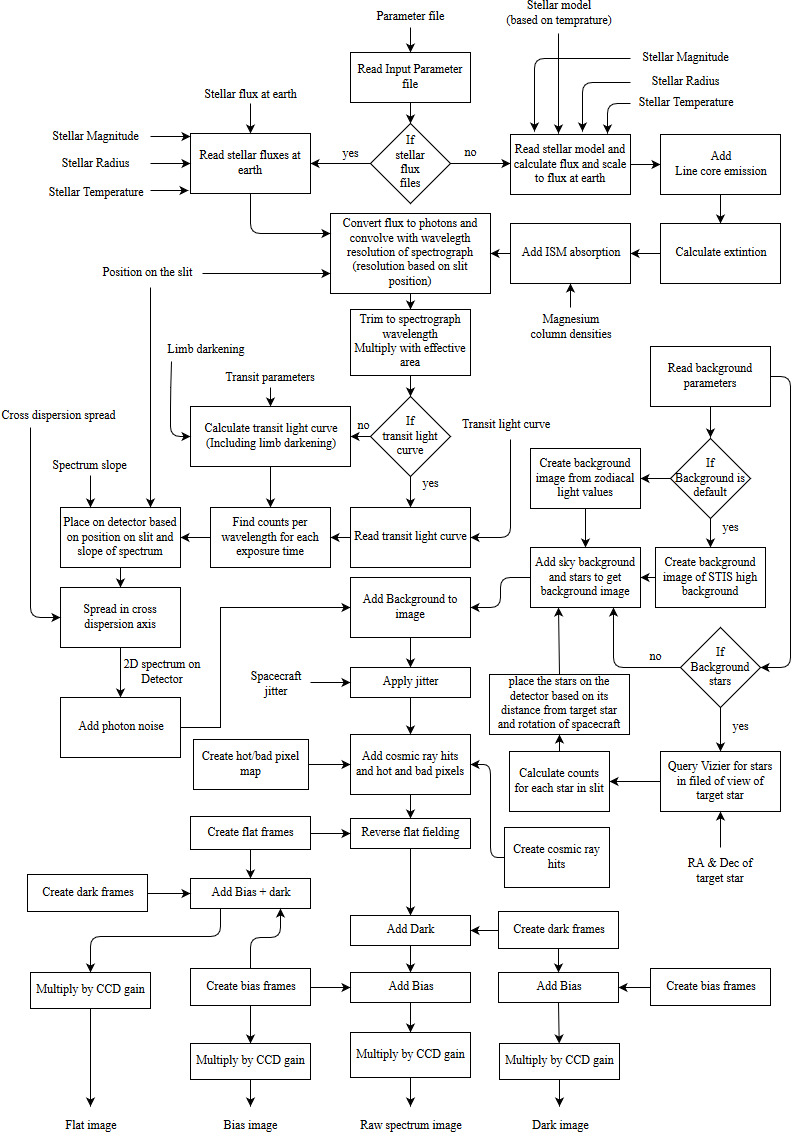}
\caption{Flowchart of the CUTE data simulator.}
\label{fig:fig2}
\end{center}
\end{figure}

\begin{figure}[h!]
\begin{center}
\includegraphics[width=\textwidth]{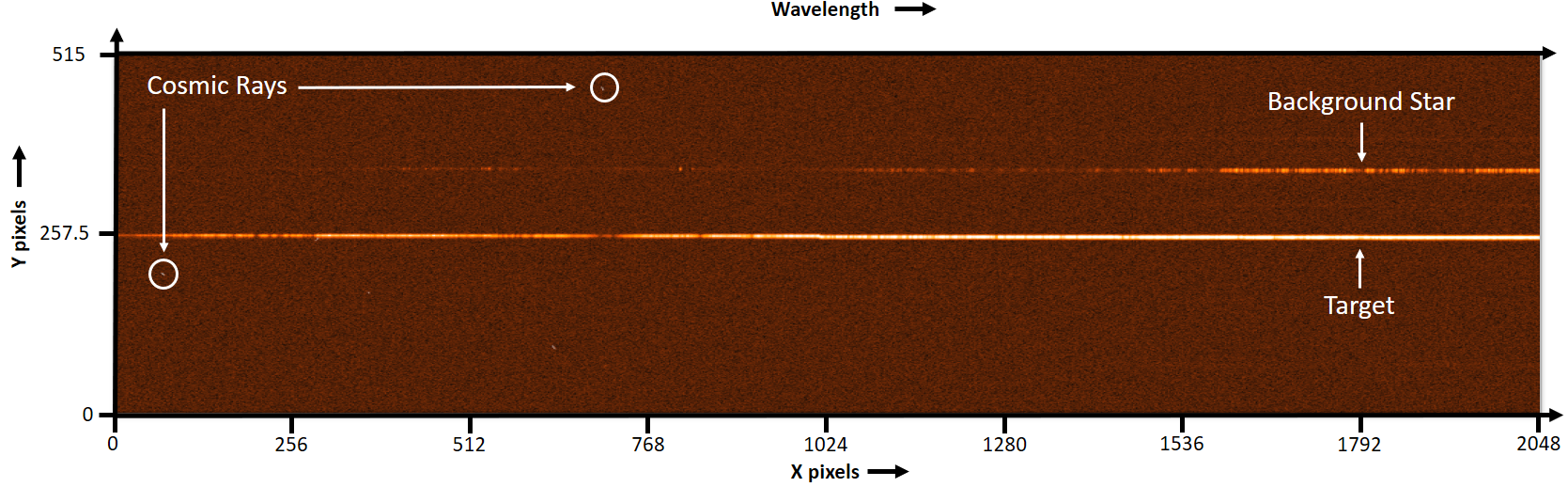}
\caption{Example of an image generated for KELT-7 field with the simulator for CUTE instrument.}
\label{fig:fig3}
\end{center}
\end{figure}

The simulator, called ACUTEDIRNDL\footnote{The simulator is available for download at {\tt https://github.com/agsreejith/ACUTEDIRNDL}.}, is a collection of IDL routines recreating the effects of the CCD and readout electronics, the optical effects of spectrograph, the spacecraft attitude orientation and jitter effects, and background sources on an input stellar spectral energy distribution modulated with any given transit light curve. Figure~\ref{fig:fig2} shows the flowchart describing how the different routines are run in order to reach the final products and how they interact with each other. Figure~\ref{fig:fig3} shows an example output image generated with the simulator.

The software is governed by a file containing all the input parameters, both astrophysical and instrumental. Here below, we provide a brief description of the input parameters of each module. The details of the simulator architecture and of the algorithms inside each module will be discussed in a forthcoming paper. 
\subsection{Input parameters} 
\noindent 1. Stellar parameters. The relevant stellar parameters consist of the effective temperature, V-band magnitude, and radius. Based on the stellar effective temperature suitable stellar spectrum is taken from a database of pre-computed spectral energy distributions, computed with LLmodels \citep{llmodels}. The correct spectral intensity as a function of distance from the stellar centre, computed with the PHOENIX stellar atmosphere code \citep[][{\tt http://phoenix.astro.physik.uni-goettingen.de/}]{phoenix} is also selected based on stellar effective temperatures, which is used to compute the limb darkening coefficients employed to calculate the transit light curve. The simulator uses an internal look-up table roughly relating stellar spectral types, effective temperatures, $B-V$ colors, and radii. Further input of the data simulator relevant to the computation of the final stellar spectrum is the $\log{R'_{\rm HK}}$ value, which is used to compute the Mg{\sc ii}\,h\&k line core emission from the Ca{\sc ii}-to-Mg{\sc ii} emission scaling relations \citep{linsky2013} and the interstellar extinction $E(B-V)$. This last parameter is used to compute the Mg{\sc ii} column density in the interstellar medium (ISM) using relations based on $E(B-V)$, the H{\sc i} column density, and the Mg{\sc ii} ISM absorption \citep{savage1979,frisch2003}.

\noindent 2. Transit parameters. These parameters are required for the computation of the planetary transit light curve. These parameters correspond to the planet impact parameter in units of stellar radius, the planetary orbital period in days, the orbital inclination axis in degrees, the orbital semi-major axis in units of stellar radii, the planetary radius in units of stellar radius, and the mid-transit time in JD. The limb darkening coefficients are automatically computed on the basis of the PHOENIX stellar intensities.

\noindent 3. Target coordinates, namely RA and Dec, in degrees and exposure time for each observation in seconds.	

\noindent 4. Instrument parameters. These are the position of the star on the slit, the spectrograph spectral resolution in Angstroms and the shape of the spectral footprint in the cross-dispersion direction. 

\noindent 5. Background parameters. These are the level of cosmic ray hits, the level of zodiacal light, and the position, magnitude, and spectral type of the background stars. 

\noindent 6. CCD parameters. These are physical size, pixel size, gain, and read-out noise of the CCD.

\noindent 7. Spacecraft parameters. These are the parameters characterizing the spacecraft attitude in the form of jitter and orientation.

The simulator is separated into five different modules, which we describe here below.

\noindent {\bf Stellar module.} This module generates the stellar spectra energy distribution, including the effects of Mg{\sc ii} line core emission and ISM absorption.

\noindent {\bf Background module.} This module connects with ViZieR to extract the position, $B-V$ color, and V-band magnitude of the stars lying close to the target and within the CUTE field of view. The simulator uses the $B-V$ color to estimate the stellar effective temperature. This module adds also the contribution of the zodiacal light to the background. 

\noindent{\bf Transit module.} This module computes wavelength dependent synthetic light curves on the basis of the input parameters given by the user. The module calculates the transit duration on the basis of the input parameters and then uses EXOFAST \citep{exofast} to compute the transit light curve.

\noindent {\bf CCD module.} This module creates the spectrum on the detector for a specified number of x and y pixels. It also spreads the spectrum in the cross dispersion axis based on the line spread function obtained from the optical model of CUTE computed with ZEMAX.  

\noindent {\bf Spacecraft module.} This module simulates the effect of spacecraft jitter and orientation on the output image.
%
\section{Data Reduction Pipeline}
%
\begin{figure}[h!]
\begin{center}
\includegraphics[height=20.5cm]{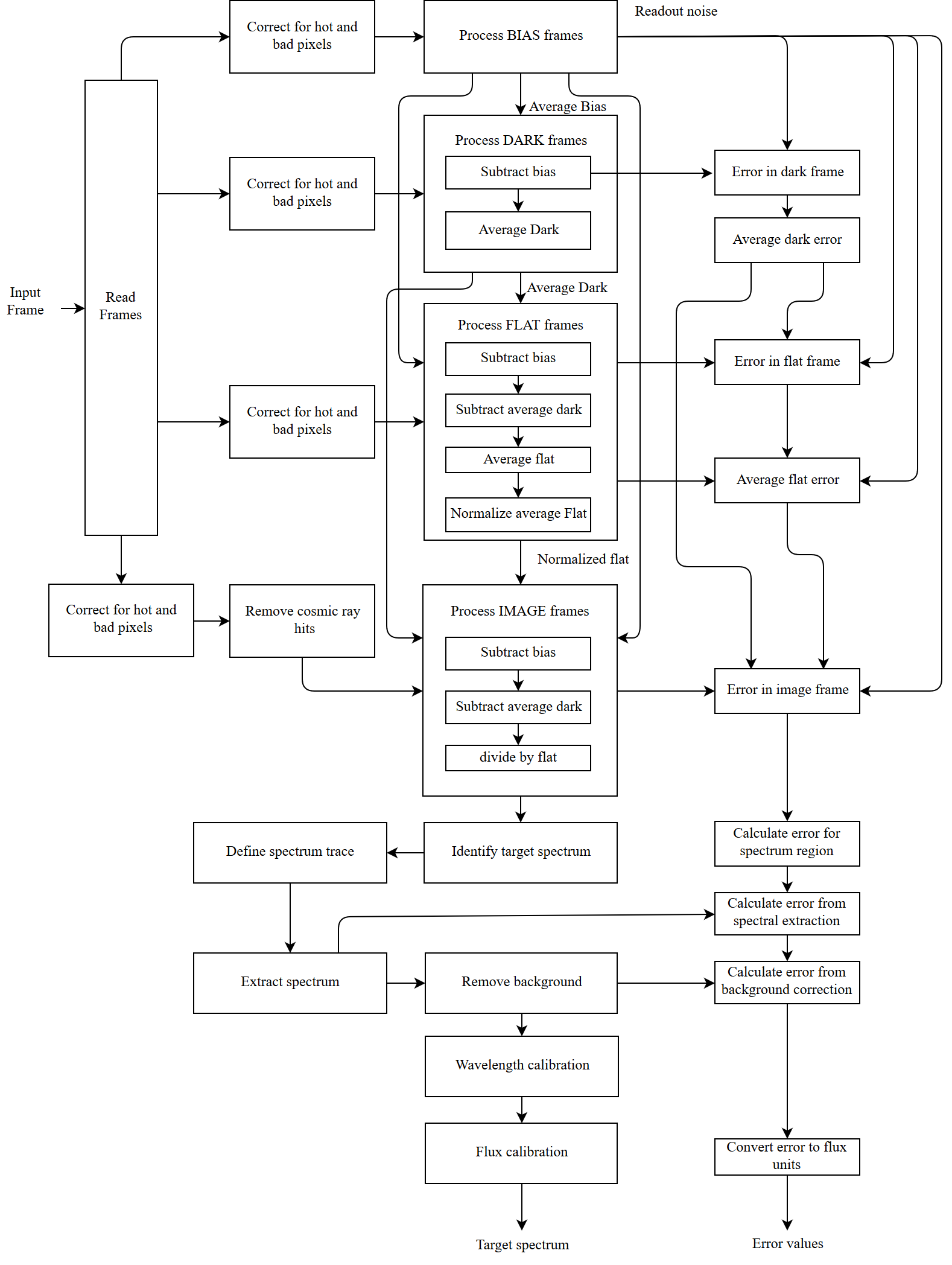}
\caption{Flowchart of the data reduction pipeline. The ground segment of the pipeline is enclosed in a dashed box.}
\label{fig:fig4}
\end{center}
\end{figure}

CUTE's orbital period around the Earth will be about 90 minutes, thus leaving a limited amount of time for communications with the ground stations in Boulder, Colorado, and Fairbanks, Alaska (USA). CUTE will employ an S-band communication between the satellite and the ground station, which would allow us to download from the satellite between 20\% - 60\% of the science data collected (depending on observing cadence), which can  then be reduced on the ground. As a primary uplink channel and backup downlink capability, CUTE has a UHF transmitter/receiver on board to communicate with the LASP ground stations.  In  the event that the UHF transmitter is our primary downlink  channel, we will employ on-board processing and transmit to the ground reduced one-dimensional spectra, housekeeping data, and several full-frame images a week to verify on-board processing and target centering.

The main tasks of the ground data reduction pipeline will be dark and bias subtraction, corrections for bad/hot pixels, cosmic-ray correction, flat-field removal, spectral extraction, background subtraction, wavelength calibration, and flux calibration. For each science exposure, the output of the pipeline will then be a wavelength and flux calibrated one-dimensional spectrum of the target. Since master dark and bias frames will be assembled on board, to be then routinely transfered to the ground to monitor the aging of the CCD (i.e., detector cosmetics and background levels), the data reduction pipeline will need to be able also to combine dark and bias frames to produce the master frames. Cosmic ray rejection will be performed using contiguous science frames \citep{dokkum}. Flat-fields will be obtained on the ground during instrument testing and the master flat-field will be stored on-board \citep{fleming} before launch and used for calibrations.

The spectral extraction will be performed by collapsing the signal in the cross-dispersion direction, taking into account the spectrum's width and shape. The exact algorithm for spectral extraction is however still to be identified. Simultaneously, the pipeline will extract a spectrum in a region of the CCD not illuminated by the target or background stars to measure the background level. Care will be taken in the preparation of the observations to minimise contamination of the target by the light of nearby stars. One of the tasks of the data simulator described above is to identify the best orientation of the spacecraft for the observations in order to minimise the contamination by background stars.

Wavelength calibration frames will be obtained on the ground, during calibration and testing, and will be used to perform the wavelength calibration of the one-dimensional spectra. During commissioning, we will observe a few nearby sharp-lined late-type stars, thus rich in spectral lines, which spectra can be used to check the quality of the ground based wavelength calibration and adjust it, if necessary. CUTE's spectral resolution is also high enough to allow the application of cross-correlation techniques to precisely measure the absolute wavelength calibration of these stars with rich spectra \citep{bagnulo}. During commissioning, we will also observe standard stars (i.e., white dwarfs) to enable flux calibration. Wavelength and flux calibration will be done on the ground.

Several full two-dimensional spectral images will be downlinked every week to monitor the spectral performance and extraction region algorithms. Extraction algorithm errors detected from these two-dimensional images will be corrected by re-reducing the raw data on the ground. This is enabled by the availability of approximately 6 GB of non-volatile storage on the XB1 spacecraft bus. A double buffered non--volatile storage scheme will be employed to always allow access to approximately 10 days of image data. 

The data reduction pipeline will be flexible enough to be able to account for the effects of possible in-flight complications. The flowchart of the pipeline is as shown in Fig.~\ref{fig:fig4}. This flowchart forms the basis of the logic of the data reduction pipeline and will be expanded upon. For a quick analysis of the data produced by the simulator, we developed a preliminary version of the data reduction pipeline, which is capable of applying all reduction steps described above, except for cosmic-ray rejection, subtraction of the background stars, and flux calibration.
%

\section{Summary and conclusions}
We presented the CUTE data simulator, which is capable of simulating near-ultraviolet transmission spectroscopy. The CUTE data simulator is complemented by a reduction pipeline enabling a rough reduction and analysis of the simulated spectra. The simulator is constantly being updated to improve its performances, fix minor bugs, and adjust for changes in the CUTE telescope and instrument design and/or in the mission operations. The simulator has been designed and developed in such a way that it can be easily adapted to any other mission performing single-slit spectroscopy and carrying on-board a CCD detector.

The main aim of the data simulator is to test the observability of the selected targets and the quality of the data. This is a particularly important aspect given that the NASA TESS mission \citep{tess} is expected to find thousands of transiting exoplanets, dozens of them amenable to CUTE observations. It will also be used to refine the signal-to-noise calculator and to decide some of the spacecraft and instrument settings (e.g., slit orientation, exposure times, position of the target along the slit) prior to observations. The simulator will become important also to study the effect on the data of deviations from the nominal instrument configurations that may become evident during CUTE's assembling and testing, or after launch. More importantly, this will allow us to prepare a data reduction pipeline capable of dealing with data that may be affected by such deviations. This approach reduces mission risk and increases operational efficiency. The light curves generated by the simulator will be also used to train the science team in the analysis of the CUTE data, prior to launch. 


\acknowledgments 

 A.~G.~Sreejith, L. Fossati, and M. Stellar acknowledge financial support from the Austrian\\ Forschungsf\"orderungsgesellschaft FFG project “ACUTEDIRNDL” P859718. CUTE is supported by NASA grant NNX17AI84G (PI - K. France) to the University of Colorado Boulder.  

\bibliography{report} 
\bibliographystyle{spiebib} 

\end{document}